\PassOptionsToPackage{table,xcdraw}{xcolor}

\documentclass[sigconf,screen,authorversion]{acmart}

\AtBeginDocument{%
  }

\copyrightyear{2026}
\acmYear{2026}
\setcopyright{cc}
\setcctype{by}
\acmConference[LAK 2026]{LAK26: 16th International Learning Analytics and Knowledge Conference}{April 27-May 01, 2026}{Bergen, Norway}
\acmBooktitle{LAK26: 16th International Learning Analytics and Knowledge Conference (LAK 2026), April 27-May 01, 2026, Bergen, Norway}
\acmPrice{}
\acmDOI{10.1145/3785022.3785025}
\acmISBN{979-8-4007-2066-6/2026/04}

\begin{document}

\title{Understanding Gaming the System by Analyzing Self-Regulated Learning in Think-Aloud Protocols}

\author{Jiayi Zhang}
\orcid{0000-0002-7334-4256}
\affiliation{
  \institution{University of Pennsylvania}
  \city{Philadelphia, PA}
  \country{USA}
}
\email{joycez@upenn.edu}

\author{Conrad Borchers}
\orcid{0000-0003-3437-8979}
\affiliation{
  \institution{Carnegie Mellon University}
\city{Pittsburgh, PA}
  \country{USA}
}
\email{cborcher@cs.cmu.edu}

\author{Canwen Wang}
\orcid{0009-0002-1434-224X}
\affiliation{
  \institution{Carnegie Mellon University}
\city{Pittsburgh, PA}
  \country{USA}
}
\email{canwenw@andrew.cmu.edu}

\author{Vishal Kumar}
\orcid{0009-0001-6213-0033}
\affiliation{
  \institution{Stanford University}
  \city{Stanford, CA}
  \country{USA}
}
\email{vishalkumarswe@gmail.com}

\author{Leah Teffera}
\orcid{0009-0002-0886-4412}
\affiliation{
  \institution{Carnegie Mellon University}
\city{Pittsburgh, PA}
  \country{USA}
}
\email{lteffera@andrew.cmu.edu}

\author{Bruce M. McLaren}
\orcid{0000-0002-1196-5284}
\affiliation{
  \institution{Carnegie Mellon University}
\city{Pittsburgh, PA}
  \country{USA}
}
\email{bmclaren@andrew.cmu.edu}

\author{Ryan S. Baker}
\orcid{0000-0002-3051-3232}
\affiliation{
  \institution{Adelaide University}
  \city{Adelaide, SA}
  \country{Australia}
}
\email{ryanshaunbaker@gmail.com}

\makeatletter
\let\@authorsaddresses\@empty
\makeatother

\renewcommand{\shortauthors}{Zhang et al.}

\begin{abstract}
In digital learning systems, gaming the system refers to occasions when students attempt to succeed in an educational task by systematically taking advantage of system features rather than engaging meaningfully with the content. Often viewed as a form of behavioral disengagement, gaming the system is negatively associated with short- and long-term learning outcomes. However, little research has explored this phenomenon beyond its behavioral representation, leaving questions such as whether students are cognitively disengaged or whether they engage in different self-regulated learning (SRL) strategies when gaming largely unanswered. This study employs a mixed-methods approach to examine students’ cognitive engagement and SRL processes during gaming versus non-gaming periods, using utterance length and SRL codes inferred from think-aloud protocols collected while students interacted with an intelligent tutoring system for chemistry. We found that gaming does not simply reflect a lack of cognitive effort; during gaming, students often produced longer utterances, were more likely to engage in processing information and realizing errors, but less likely to engage in planning, and exhibited reactive rather than proactive self-regulatory strategies. These findings provide empirical evidence supporting the interpretation that gaming may represent a maladaptive form of SRL. With this understanding, future work can address gaming and its negative impacts by designing systems that target maladaptive self-regulation to promote better learning.
\end{abstract}



\maketitle

\section{Introduction}
\label{sec:intro}
Many contemporary AI-based learning systems, such as tutoring systems or digital learning games, provide learners with adaptive support to complete problem-solving steps when they need it. However, assistance in learning systems can also be exploited, much to the detriment of learning. One extensively studied phenomenon to describe learners exploiting assistance is gaming the system \cite{baker2004offtask}. Gaming the system, often framed as a form of behavioral disengagement, occurs when learners attempt to succeed in educational tasks by systematically exploiting the system’s properties and regularities, rather than by thinking through the material \cite{baker2004offtask}. In such cases, students may engage in systematic guessing or hint abuse, completing an educational task without learning the materials. Understanding this behavior is critical: by capturing not only when students exploit system features but also the underlying mechanisms that drive such behavior, researchers and designers can develop interventions that reduce gaming and foster deeper and more productive learning.

Often construed as a form of behavioral disengagement, detecting gaming involves the use of behavioral logs, where rules \cite{muldner2010gaming} or machine-learned models \cite{baker2010constraint,richey2021gaming, zhang2025llm} are applied to identify behavioral patterns, such as systematic guessing where students make rapid guesses in order to get the correct answer. Previous studies have repeatedly found that gaming the system is negatively correlated with both short- and long-term learning outcomes, showing that students who had higher rates of gaming were associated with lower grades, less learning, and were less likely to enroll in STEM-related majors \cite{baker2004offtask,richey2021gaming,almeda2020stem,fancsali2015carelessness,pardos2014affective,baker2025gaming, zhang2025llm}. 

While numerous studies have investigated this behavior, including understanding the affect and motivations associated with the choice to game \cite{fancsali2015carelessness,baker2010frustrated,muldner2010gaming}, research has not yet fully explored the cognition that occurs when students game the system. As a result, our understanding of the cognitive processes that occur during gaming behaviors remains limited, leaving questions, such as whether students are also cognitively disengaged when they are behaviorally disengaged, or whether they engage in different self-regulated strategies when gaming the system, largely unanswered. Without a clear understanding of the mechanisms driving gaming behavior, interventions to address gaming may be less effective than expected, and we may not be able to predict which interventions will and will not work. 

To this end, Baker and colleagues attempt to explain gaming the system beyond behavioral representation, theorizing that gaming the system may be an extreme form of self-regulated learning (SRL) \cite{baker2013handbook}. Specifically, they posit that gaming the system can be viewed as “self-regulated non-learning”, hence a maladaptive use of SRL. In this context, a student who games the system is typically not attempting to learn at all during their gaming behavior. As a result, this intention of “not attempting to learn” may lead to a different SRL use and various levels or types of cognitive engagement. While the relationship between SRL and gaming has been suggested, no studies have empirically demonstrated this hypothesized connection between gaming the system (a behavioral disengagement) and SRL (a multidimensional self-guided learning process), illustrating the differential SRL use between scenarios when students game the system and when they do not. 

Understanding the relationship between gaming and SRL has both theoretical and practical implications. Theoretically, it could refine models of SRL by integrating maladaptive or context-contingent strategies, as most SRL theories are based on a gold standard theoretical cycle of how SRL should ideally be carried out \cite{panadero2017review,winne1998srl,zimmerman2000selfreg}. Practically, it may explain why some forms of gaming interventions (e.g., discouragement messages) have shown success in reducing gaming behavior \cite{baker2006adapting}, while others (e.g., gamification) have shown limited success \cite{azevedo2018gamification}. If gaming is indeed a form of strategic regulation, then interventions may need to shift from addressing behaviors to modifying the learning environment and providing SRL support, such as goal setting, metacognitive prompts, and progress monitoring \cite{arroyo2007repairing,xia2020viz}. However, to arrive at such interventions, we must better understand how gaming relates to SRL and when it may be considered a beneficial or maladaptive SRL strategy.

Therefore, in this study, we investigated students’ cognitive engagement and SRL behaviors (inferred from think-aloud utterances) in relation to gaming the system during students' interactions with an intelligent tutoring system. A mixed-method approach was employed to analyze both the quantitative and qualitative aspects of the utterances, providing insights into students' cognitive engagement and SRL use during gaming and non-gaming scenarios. Specifically, for \textbf{RQ1}, using the length of utterances as an indicator of cognitive engagement (as proposed and used in \cite{li2022cogengagement}), we evaluated students’ cognitive engagement in gaming and non-gaming scenarios. We examined whether students' utterances were systematically longer or shorter during the time periods when gaming the system was observed. This analysis helps to determine whether gaming is associated with reduced verbal expression, which may signal lower levels of cognitive engagement. For the qualitative analysis, we labeled four SRL categories (\textit{Processing Information, Planning, Enacting, and Realizing Errors}) evident in students’ utterances and analyzed differences in their presence and transitions between the gaming and non-gaming scenarios. Specifically, \textbf{RQ2} explored whether students were more likely to engage in any of the four SRL categories when gaming, providing insights on which specific regulatory processes may be suppressed or maintained during this disengaged behavior. Given that SRL is temporal and dynamic in nature \cite{panadero2017review}, through using ordered network analysis, \textbf{RQ3} examined differences in SRL transitions, providing a more fine-grained comparison that captures the dynamic process of self-regulatory behaviors. Specifically, the ordered networks visualize the directed connections among SRL categories, highlighting differences in students’ use of self-regulated learning strategies during gaming and non-gaming scenarios. 

Through these analyses, we aim to gain a deeper understanding of the cognitive and self-regulatory processes underlying gaming behavior. This work advances the theoretical understanding of the cognition surrounding gaming the system. Such insights can inform the design of interventions that more effectively address the issue and mitigate its negative impact. Additionally, this study demonstrates the use and integration of log data and think-aloud utterances, a data combination less often seen in learning analytics research, showing how combining these sources can enrich the conceptual understanding of complex learning behaviors.

\section{Related Literature}
\label{sec: related literature}

\subsection{Gaming the System}
\label{sec: gaming the system}
Gaming the system refers to student behaviors aimed at exploiting the structure of learning systems to progress without engaging meaningfully with the learning content. Such behaviors are associated with little to no learning \cite{gong2010gaming}. In the field of learning analytics, gaming is typically measured through log-based detection of behaviors such as rapid guessing, or hint abuse \cite{baker2004offtask}. This line of work has led to many successful applications of gaming detection to grade prediction \cite{pardos2014affective}, dropout prediction \cite{Karumbaiah2018PredictingQuitting}, and analytics that help guide teacher attention toward effective instructional interventions \cite{holstein2019codesigning}.

The phenomenon of gaming the system has been observed since the very beginning of computer-based tutoring, with early research from the 1970s already noting that some students quickly used correct answers provided in feedback without processing the remaining feedback in order to advance in the system \cite{tait1973feedback}. The term gaming the system was formally introduced in 2004 \cite{baker2004offtask}, alongside the hypothesis that such behavior may, in some cases, represent an adaptive strategy, particularly among students with a performance orientation who prioritize task completion over deep understanding.

One challenge in the measurement of gaming the system is that the relevant learner behaviors are known to differ by learning context. In fact, gaming appears to be substantially influenced by system design rather than student characteristics. For example, one study found that interface features, such as poorly designed toolbars or abstract hint formats, accounted for up to 56\% of the variance in gaming behavior, a figure more than five times greater than variance explained by student-level traits such as interest in mathematics \cite{baker2009software}. Similarly, certain content-level factors, including excessive difficulty or lack of instructional clarity, have been shown to provoke gaming behavior \cite{baker2009software,slater2016semantic}. Researchers have adjusted gaming detection models to such context differences to assess student-level gaming tendencies and better understand learning \cite{huang2023using}.

This body of evidence supports the view that gaming is not only a student deficit (i.e., a student generally being unwilling to learn) but also a context-sensitive response—a behavioral adaptation to the perceived demands and affordances of the learning environment. This perspective aligns with broader theories of SRL, which emphasize how learners strategically manage their cognition, motivation, and behavior in response to task demands \cite{zhang2024srlcycles}. And indeed, at least one study suggests that high levels of SRL are negatively correlated with behaviors indicative of gaming \cite{heirweg2019profiling}. This leads to an understudied hypothesis: gaming the system may represent a form of self-regulated learning, albeit one that is suboptimal or misaligned with learning goals. 

\subsection{Self-Regulated Behaviors and Learning}
\label{sec: self-regulated behaviors and learning }
Self-regulated learning (SRL) describes the process in which students direct their attention and effort in pursuit of goals during learning \cite{zimmerman2000selfreg}. A range of cognitive, affective, metacognitive, and motivational processes are involved in SRL \cite{panadero2017review}. Despite differences in emphases, most foundational theories characterize SRL as a cyclical process consisting of phases in which learners define tasks, make plans, enact those plans, and reflect and adapt \cite{panadero2017review,zimmerman2000selfreg,winne1998srl}. For example, grounded in information processing theory, Winne and Hadwin \cite{winne1998srl} describe SRL as four interdependent and cyclical stages, including (1) task definition, (2) goal setting and planning, (3) plan enactment, and (4) reflection and adaptation when goals are not met. Across these stages, learners employ cognitive and metacognitive strategies, such as monitoring and reflecting, to accomplish the task.

Previous studies have consistently found a positive relationship between the use of SRL and academic achievement \cite{zimmerman2013theories,viberg2020self}. For instance, students who regularly engage in self-regulatory practices such as planning and reflecting tend to be more successful in learning \cite{heirweg2019profiling,bannert2014process}. Given the temporal and cyclical nature of SRL, recent research has moved beyond simply examining the presence and frequency of SRL use to analyzing SRL behaviors and processes as sequences of events, offering deeper insights into the strategic use of SRL \cite{bannert2014process,heirweg2019profiling}. For example, using process mining, \cite{heirweg2019profiling} showed distinct patterns of SRL use between high- and low-achieving students, demonstrating that high achievers engaged in a more coherent and strategic use of SRL.

Since then, network analysis methods, such as epistemic and ordered network analysis, have been adopted in numerous recent papers to examine the SRL process by analyzing the co-occurrences and transitions among SRL behaviors \cite{li2020temporal,cheng2025self,sung2024beyond, huang2023cooccurrence}. These approaches help visualize how SRL behaviors are being engaged in a pattern, and how the patterns vary across different student populations or learning contexts. For example, \cite{bannert2014process} used epistemic network analysis to compare SRL behaviors like reading and revising between high- and low-performing students in open-ended problem-solving tasks, and found that high performers exhibited patterns more consistent with established SRL cycles. Relatedly, \cite{li2020temporal} constructed ordered networks for students classified as unsuccessful, successful, or mastery-oriented. Results indicated that mastery-oriented students exhibited denser networks and behavior sequences aligned with theoretical SRL models, where early strategies influenced subsequent actions. Other research has shown that SRL behavior patterns shift in response to task difficulty \cite{huang2023cooccurrence}, platform design \cite{zhang2024srlcycles}, or as students become more accustomed to a digital learning environment \cite{wu2022smart}. Work in this area collectively demonstrates the effectiveness of network analysis as an approach to examining SRL as a complex and temporal learning process, and highlights the positive impact of SRL when it is strategically applied in alignment with the SRL cycle proposed in foundational theories.

While previous research highlights the benefits of effective SRL, students may also, whether inadvertently or strategically, engage in ineffective or maladaptive forms of regulation \cite{baumeister1996self}. In other words, self-regulatory behaviors are not inherently beneficial; their effectiveness depends on how they are enacted and the goals they intend to serve. In some cases, students may process information superficially \cite{winne2018theorizing}, enact plans without reflection \cite{li2020temporal}, or rely on strategies that do not advance learning or even strategies intentionally chosen to avoid learning \cite{skaalvik2018mathematics}. Such behaviors are regulatory in nature, but do not lead to the positive outcomes we hope learners will achieve \cite{baumeister1996self}.

Building on this idea, \cite{baker2013handbook} asked whether gaming the system can be understood as an ineffective form of SRL. From this perspective, gaming is not simply a disengaged or off-task behavior but rather a self-initiated regulatory strategy (albeit a maladaptive one) that students adopt in response to external factors (e.g., system design, scaffolding, task difficulty) and internal factors (e.g., motivation, frustration, prior failure). Reconceptualizing gaming in this way enriches its definition: rather than viewing it solely as problematic disengaged behavior, it may also represent a strategic, though counterproductive, choice made by students in particular contexts. This framing carries important implications. If gaming reflects a maladaptive use of SRL, interventions should not be limited to discouraging or penalizing the behavior, or towards making it more difficult to engage in \cite{li2022multi,baker2006adapting,walonoski2006prevention}. Instead, learning systems can be designed to identify and address the underlying regulatory mechanisms that lead to gaming, and to scaffold students toward more adaptive and productive use of SRL \cite{arroyo2007repairing}. In this way, learning analytics can move beyond simply responding to disengagement to supporting the use of effective self-regulation.

\section{Methods}
\label{sec:methods}
In this study, we investigated gaming the system using a multimodal dataset that included log data and think-aloud utterances from students solving chemistry problems with the Stoichiometry Tutor (Section \ref{sec:learningplatforms}). While solving problems, students were instructed to verbalize their thoughts (Section \ref{sec:participants}). Two researchers coded the utterances for four SRL categories (\textit{Processing Information, Planning, Enacting, and Realizing Errors}) aligned with Winne and Hadwin’s four-stage model \cite{winne1998srl} (Section \ref{sec:code srl}). Simultaneously, log data was segmented into clips, and each was coded for the presence or absence of gaming the system using text replay coding (Section \ref{sec:code gaming}). We combined the two data sources, think-aloud data (along with the SRL codes) and log data (Section \ref{sec:analysis data set}), and conducted a series of analyses comparing utterances and SRL engagement between gaming and non-gaming clips to answer the three research questions.

\subsection{Learning platforms}
\label{sec:learningplatforms}

\begin{figure}[htp]
    \centering
    \includegraphics[width=\linewidth]{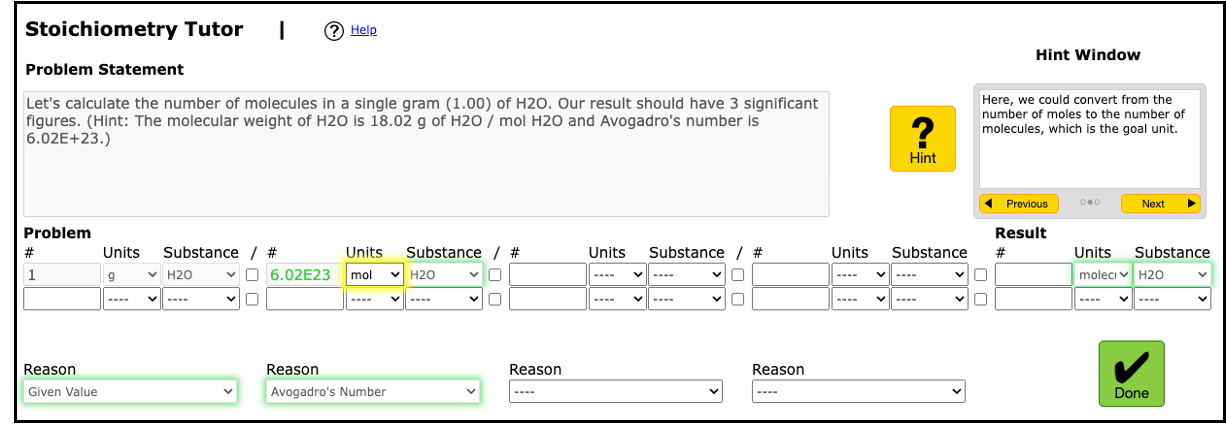}
    \caption{Interface example of StoichTutor}
    \label{fig:stoich}
\end{figure}

StoichTutor (see Fig. \ref{fig:stoich}) is an intelligent tutoring system designed to teach stoichiometry using the factor-label method, guiding learners to systematically convert units from a given to a target value \cite{mclaren2016worked}. StoichTutor requires learners to enter values or select values from drop-down menus, and it provides immediate feedback and hints. Students must complete all steps to advance to the next problem. StoichTutor has been shown to significantly improve students' stoichiometry skills \cite{mclaren2016worked,borchers2025scaffolding}.

\subsection{Participants}
\label{sec:participants}
Ten students, comprising 9 undergraduates and 1 graduate student, participated in a study conducted in 2023 \cite{borchers2025stoichiometry}. 
All participants were enrolled in U.S. degree programs, with self-reported prior proficiency in stoichiometry averaging 3.4 (SD = 1.35) on a five-point Likert scale. Recruitment involved course announcements and snowball sampling in relevant classes. Each participant received a \$15 Amazon gift card. 

Students were instructed to solve stoichiometry problems using StoichTutor and engage in a think-aloud protocol, verbalizing their thinking process while solving the problem \cite{borchers2024using}.

In accordance with standard practice, if a student remained silent for more than five seconds, a researcher prompted them to continue speaking. These verbalizations were recorded and transcribed using Whisper, an open-source transcription model that segmented the utterances with start and end timestamps. A comprehensive report on Whisper’s accuracy and error types is given in \cite{radford2023robust}.

\subsection{Coding SRL Categories in the Think-Aloud Data}
\label{sec:code srl}
To analyze the think-aloud data, we merged utterances falling between two consecutive time-stamped student transactions/ actions, such as making attempts or requesting hints. This allowed us to analyze the utterances in a longer time window to infer students’ use of SRL strategies, as done in \cite{borchers2024using}. 

Utterances were annotated using a coding scheme (see Table~\ref{tab:codebook}), developed based on the four-stage model by Winne and Hadwin \cite{winne1998srl}. This model describes the SRL process as four interdependent and recursive stages in which learners: (1) define the task, (2) set goals and form plans, (3) enact the plans, and (4) reflect and adapt strategies when goals are not met. To capture students’ engagement in SRL at each stage, we identified key behaviors that are both relevant and salient (i.e., observable) in students’ utterances. We operationalized these key behaviors into four SRL categories: \textit{Processing Information}, \textit{Planning}, \textit{Enacting}, and \textit{Realizing Errors}. These categories, focusing on relevant behaviors within problem-solving learning environments, represent a subset of SRL behaviors within each stage of the four-stage model. Table~\ref{tab:srl} outlines the coding categories, indicative behaviors, and example utterances.

Two coders established acceptable inter-rater reliability after coding 162 utterances 
($\kappa_{\text{processing}} = 0.78$, 
$\kappa_{\text{planning}} = 0.90$, 
$\kappa_{\text{enacting}} = 0.77$, 
$\kappa_{\text{errors}} = 1.00$). 
They then individually coded the remaining utterances. In total, students generated 401 utterances while using StoichTutor. Among them, 15\% were coded as \textit{Processing Information}, 12\% as \textit{Planning}, 25\% as \textit{Enacting}, and 7\% as \textit{Realizing Errors}.

\begin{table*}[htp]
\captionof{table}{Overview of the four SRL labels, including indicative behaviors of each label and example utterances.}
\label{tab:codebook}
\centering
\begin{tabular}{|p{0.12\textwidth}|p{0.545\textwidth}|p{0.265\textwidth}|}
\hline
SRL Category & Behaviors & Example Utterance \\
\hline
Processing\newline Information & Assemble information: \begin{itemize}
    \item The utterance demonstrates behaviors where students read or re-read a question, hints, or feedback provided by the system
\end{itemize}
Comprehend information:
\begin{itemize}
\item The utterance demonstrates behaviors where students repeat information provided by the system with a level of synthesis
\end{itemize}
                     & \textit{``Let's figure out how many hydrogen items are in a millimole of water molecule H2O molecules. Our result should have three significant features. Figures. Avogadro number is 6.02 E23. 2 atoms of H2O.''} \\
\hline
Planning & Identify goals and form plans:
\begin{itemize}
    \item The utterance reflects behaviors where students verbalize a conceptual plan of how they will solve the problem 
\end{itemize}
            & \textit{``Our goal of the result is hydrogen atoms. The goal of the result is the number of hydrogen atoms, right?''} \\
\hline
Enacting & 
Verbalize previous action:
\begin{itemize}
    \item The utterance reflects students' behaviors where they verbalize an action that has just been carried out, explaining what they did
\end{itemize}
Announce the next action:
\begin{itemize}
    \item The utterance reflects student behaviors where they verbalize a concrete and specific action that they will do next
\end{itemize}
             & \textit{``Two molecules of this. How many atoms in a... How many atoms in a minimum molecule of M mole? 61023 divided by 2. 3.0115.''} \\
\hline
Realizing\\Errors & 
Realize something is wrong:
\begin{itemize}
    \item The utterance demonstrates instances where students realize there is a mistake in the answer or the process, with or without external prompting (i.e., tutor feedback)
\end{itemize}
                     & \textit{``It's incorrect. What's happened? It is the thousand in the wrong spot. 32 grams per mole. No, the thousand is correct, so what am I doing wrong? [...]''} \\
\hline
\end{tabular}
\end{table*}

\subsection{Coding Gaming the System in Log Data}
\label{sec:code gaming}
\begin{figure}[htbp]
    \centering
    \includegraphics[width=.8\linewidth]{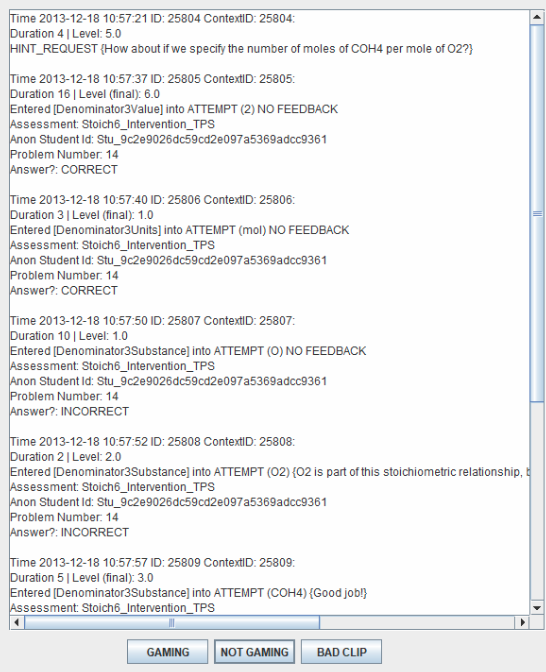}
    \caption{Text replay interface}
    \label{fig:text replay}
\end{figure}

To identify instances of gaming the system, we used text replay coding \cite{baker2008labeling} to analyze and annotate learners’ actions by reviewing system logs alongside contextual information, determining whether gaming occurred within short time periods. Specifically, we first extracted log data into clips with each clip consisting of eight transactions/actions (see Fig. \ref{fig:text replay} for an example). Each transaction reflects how a student interacts with the platform, such as clicking a hint or entering a value. All transactions are timestamped, which allows us to compute the duration of each action and link them to the corresponding timestamped utterances.

The size of the clips is in line with previous literature in determining gaming the system in ITS that employs rule-based structures \cite{baker2010constraint}. Similar to previous work, two researchers discussed and identified behavioral patterns (such as hint abuse and systematic guessing) that are indicative of gaming the system and coded the same set of clips separately to establish inter-rater reliability. Once the inter-rater reliability had been established on 100 clips
($\kappa_{\text{gaming}} = 0.86$), the coders proceeded to label the rest of the clips. In total, 117 clips were extracted. 39 clips (33\%) were coded as involving gaming the system and 78 clips were coded as not involving gaming the system, an unusually high proportion of gaming the system compared to many past studies (around 3-10\% \cite{baker2004offtask,paquette2015crosssystem}) but consistent with other tutoring systems with similar designs (e.g., rule-based structures, scaffolded platform design, and use of hints; typically around 20-40\% \cite{baker2010constraint,richey2021gaming}).

\subsection{Analysis Data Set}
\label{sec:analysis data set}
\begin{figure}[htbp]
    \centering
    \includegraphics[width=\linewidth]{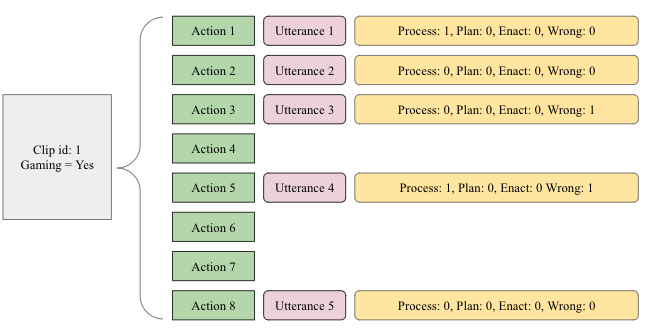}
    \caption{Data structure of a clip}
    \label{fig:clip}
\end{figure}

To understand gaming the system through think-aloud protocols, we merged think-aloud utterances (along with their SRL codes) with log data. Specifically, using timestamps, each utterance was associated with the closest preceding student transaction/action. In this way, utterances (and their corresponding SRL codes) were linked to actions in a clip. As shown in Fig. \ref{fig:clip} (the data structure of a clip), each clip contains eight actions (Actions 1–8), and each utterance is associated with an action using timestamps. If students did not speak between two consecutive actions, no utterance was associated with the former action (e.g., no utterance is linked to Action 4).

To answer RQ1 (utterance length) and RQ2 (engagement in SRL processes), we aggregated utterances and SRL categories at the clip level. For RQ1, we concatenated all utterances within a clip and counted the total number of words and the total number of non–stop words. Using generalized linear regression, we evaluated whether students were likely to produce significantly longer utterances in clips where gaming was observed. For RQ2, we created a binary variable for each SRL category to indicate the presence or absence of that category within a clip. Logistic regression was then employed to evaluate whether students were more likely to engage in the four SRL categories in clips where gaming was observed.

For RQ3, the data were analyzed at the student-action level (as shown in Fig. \ref{fig:clip}) to examine SRL transitions within gaming and non-gaming clips using Ordered Network Analysis (ONA). ONA provides a fine-grained analysis by visualizing the directed co-occurrences (transitions) of constructs or codes in a conversation. This approach has been increasingly adopted in the field of learning analytics to analyze SRL behaviors and processes, given the dynamic and temporal nature of SRL \cite{li2020temporal,sung2024beyond}. For RQ3, we used ONA to visualize the transitions between SRL categories within a clip, enabling a more detailed comparison that highlights the dynamics of SRL use in gaming versus non-gaming scenarios.

\section{Results}
\label{sec:results}

\subsection{Is There a Difference in the Length of Utterances Within Gaming and Non-Gaming Clips?}
To understand students’ cognitive engagement when they were behaviorally disengaged (i.e., gaming the system), we investigated whether students’ utterances were systematically longer or shorter during gaming. We concatenated all utterances within each clip and counted the total number of words and the total number of words excluding stop words, using the \texttt{stop\_words} dictionary from the \texttt{tidytext} package.

A generalized linear regression model with a log-linear Poisson link was used to analyze the relationship between utterance length and the presence of gaming the system (see Eq.~\ref{eq:utterance}). Given that the number of utterances is not independent within students, which may bias $p$-values of model coefficients toward significance, we added a random student intercept to each model. This adjustment accounted for each student's baseline utterance length using generalized linear mixed modeling.

\begin{equation}
N_{words} = \tau_{\text{student}} + \beta_{\text{gaming}}
\label{eq:utterance}
\end{equation}
 
As shown in Table~\ref{tab:utterances}, in the 78 non-gaming clips, the average utterance length was 74 words, while the average length with stop words removed was 20 words. In the 39 gaming clips, the average utterance length was 105 words, and the average length with stop words removed was 29 words. Results from the generalized linear regression suggest that students used significantly more words in their think-aloud utterances during periods when gaming was observed, as indicated by incidence rate ratios (IRRs) greater than 1 when comparing utterance length using all words (IRR = 1.18, $p < .001$) and when comparing utterance length with stop words removed (i.e., non-stop words; IRR = 1.21, $p < .001$).

\begin{table}[ht]
\centering
\caption{Utterance length within non-gaming and gaming clips. IRR = Incidence Rate Ratio, CI = Confidence Interval.}
\resizebox{\columnwidth}{!}{%
\renewcommand{\arraystretch}{1.2}
\begin{tabular}{|l|c|c|c|c|c|}
\hline
 & \textbf{Non-gaming} & \textbf{Gaming} & \textbf{IRR} & \textbf{95\% CI} & \textbf{\textit{p}} \\
\hline
\# of words & 74 & 105 & 1.18 & 1.13 -- 1.23 & $< .001$ \\
\# of non-stop words & 20 & 29 & 1.21 & 1.12 -- 1.32 & $< .001$ \\
\hline
\end{tabular}}
\label{tab:utterances}
\end{table}

\subsection{Is There a Difference in the Likelihood of Using SRL Within Gaming and Non-Gaming Clips?}
To evaluate whether there was a significant difference in students’ likelihood of using each type of SRL strategy between gaming and non-gaming clips, we identified whether each SRL category was present in a clip. We computed the percentage of clips in which each SRL category was present for both gaming and non-gaming clips. A logistic regression with a random intercept for students was used to analyze the relationship between the likelihood of using each SRL category and gaming the system.

As shown in Table~\ref{tab:srl}, among the 78 non-gaming clips, the presence of \textit{Processing Information}, \textit{Planning}, \textit{Enacting}, and \textit{Realizing Errors} was 24\%, 27\%, 51\%, and 19\%, respectively. In gaming clips, the corresponding percentages were 46\%, 21\%, 46\%, and 26\%. Logistic regression results indicated that students were more likely to engage in \textit{Processing Information} (OR = 3.46, $p = .021$) and \textit{Realizing Errors} (OR = 1.42, $p < .001$) in clips where gaming was observed. 

In these cases, students might utilize system information such as feedback and hints to derive an answer. For example, in a gaming clip, one student read the hint, which suggested where the number should go, as reflected in their utterance: \textit{“So the molecular weight is 32.04 grams. You may want to use grams in this term, but do you want to use it in the numerator?”}

During periods when gaming was observed, students were also more likely to enunciate errors, making utterances such as: \textit{“Oh, that’s incorrect”} or \textit{“What’s wrong with my answer?”}

Conversely, students were more likely to engage in \textit{Planning} during clips in which gaming was not observed (OR = 0.55, $p < .001$). In these instances, students announced the steps they planned to take, making utterances such as: \textit{“Okay, first we have to do the unit conversion, which is in kiloliters.”}

\begin{table}[ht]
\centering
\caption{The presence of SRL categories within non-gaming and gaming clips. OR = Odds Ratio, CI = Confidence Interval.}
\resizebox{\columnwidth}{!}{%
\renewcommand{\arraystretch}{1.2}
\begin{tabular}{|l|c|c|c|c|c|}
\hline
\textbf{SRL Category} & \textbf{Non-gaming} & \textbf{Gaming} & \textbf{OR} & \textbf{95\% CI} & \textbf{\textit{p}} \\
\hline
Processing & 24\% & 46\% & 3.46 & 1.20 -- 9.96 & .021 \\
Planning & 27\% & 21\% & 0.55 & 0.54 -- 0.55 & $< .001$ \\
Enacting & 51\% & 46\% & 0.64 & 0.26 -- 1.60 & .608 \\
Realizing Errors & 19\% & 26\% & 1.42 & 1.42 -- 1.43 & $< .001$ \\
\hline
\end{tabular}}
\label{tab:srl}
\end{table}

\subsection{Is There a Difference in the Transition of SRL Categories Within Gaming and Non-Gaming Clips?}
Since SRL is temporal in nature, the sequence of SRL processes may reveal differences in the strategies students use. To better understand these differences, we examined the transitions among the four SRL categories in gaming and non-gaming clips using Ordered Network Analysis (ONA). This approach enabled us to compare SRL transitions during periods when students gamed the system versus when they did not.

ONA identifies and quantifies directed connections among codes in a segment of data (also called a conversation) by accounting for the order of events and visualizing these connections as network models \cite{tan2023ona}. We generated ordered networks using the ENA WebTool (version 1.7.0) \cite{marquart2018ena}, visualizing differences in the transitions (directed co-occurrences) of SRL categories within gaming and non-gaming clips. In this analysis, each conversation corresponds to a clip containing SRL codes derived from the utterances. To capture the fine-grained transitions among SRL categories within a clip, we used the infinite stanza. Unlike treating the entire clip as a single sequence, this method examines overlapping subsequences, allowing us to count how often directed co-occurrences (e.g., \textit{Plan} $\rightarrow$ \textit{Enact} or \textit{Plan} $\rightarrow$ \textit{Plan}) occur throughout a clip. This provides a more detailed representation of local SRL dynamics within each clip. The line weights (\textit{lw}) in the network represent the relative frequency of each directed co-occurrence. A mean rotation was applied to maximize differences along the $x$-axis. Our model yielded Spearman co-registration correlations of $r = .82$ for the first dimension and $r = .94$ for the second, indicating a strong goodness of fit.

\begin{figure*}[htbp]
    \centering
    \includegraphics[width=1\linewidth]{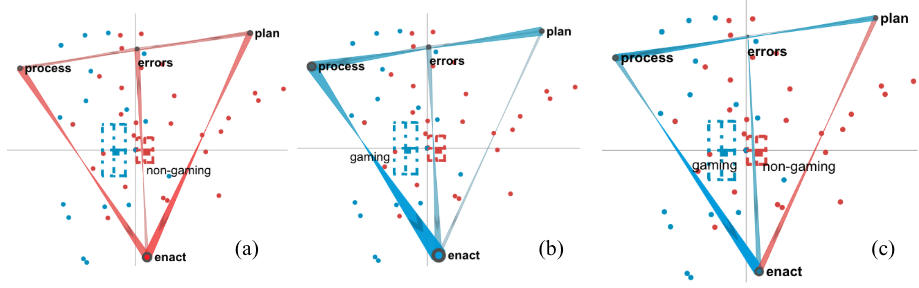}
    \caption{Ordered network of the transition of SRL categories within non-gaming (a) and gaming (b) clips, and the comparison plot (c)}
    \label{fig:ona}
\end{figure*}

Figures~\ref{fig:ona}.a and~\ref{fig:ona}.b show the ordered networks of the SRL process during the periods when students were not gaming (Fig.~\ref{fig:ona}.a) and gaming (Fig.~\ref{fig:ona}.b) the system. The line weights for each directed co-occurrence are reported in Table~\ref{tab:lineweights}. The comparison plot (Fig.~\ref{fig:ona}.c) highlights the differences in SRL transitions in the two networks between the two scenarios. 

When students were not gaming the system (Fig.~\ref{fig:ona}.a), students were likely to engage in SRL strategies aligning with the expected SRL sequence in the four-stage model \cite{winne1998srl}. Specifically, students were likely to \textit{Process Information} first and then \textit{Plan} ($lw = .05$), \textit{Plan} first and then \textit{Enact} ($lw = .09$), and \textit{Enact} first and then \textit{Realize Errors} ($lw = .06$). Additionally, both \textit{Processing Information} ($lw = .03$) and \textit{Enacting} ($lw = .03$) were likely to happen before \textit{Realizing Errors}. However, when students were gaming the system (Fig.~\ref{fig:ona}.b), the network suggests several patterns of SRL use that are misaligned with the expected SRL sequence. Specifically, students were likely to \textit{Plan} before \textit{Process Information} ($lw = .11$), and \textit{Realizing Errors} was likely to happen before \textit{Processing Information} ($lw = .05$) or before \textit{Enacting} ($lw = .08$). 

The comparison plot (Fig.~\ref{fig:ona}.c) further highlights the differences in SRL transitions in the two networks. Along the $x$-axis, a Mann-Whitney test showed that SRL processes in the non-gaming clips were significantly different from SRL processes in the gaming clips ($U = 727.50$, $p < .001$, $r = .39$). 

By comparing the two networks, we found that when students were not gaming the system (Fig.~\ref{fig:ona}.c, red), they were more likely to engage in \textit{Planning} behaviors and in the transition from \textit{Planning} to \textit{Enacting}. This transition (\textit{Plan} $\rightarrow$ \textit{Enact}) may indicate a deeper cognitive process preceding an attempt (\textit{Enact}). For example, a student said: \textit{“To get from moles to the actual number of atoms, we have to multiply by Avogadro's number [PLAN]. So, we're going to do that [...]. One thousandths of a mole times two times this per mole [ENACT].”}

In contrast, when students were gaming the system (Fig.~\ref{fig:ona}.c, blue), they were more likely to engage in \textit{Processing Information} and transition from \textit{Processing} to \textit{Enacting}. We suspect that the increased use of \textit{Processing Information} reflects scenarios in which students relied on system feedback or hints. The \textit{Process} $\rightarrow$ \textit{Enact} sequence may represent cases where students received feedback or hints (\textit{Process}) and then immediately applied them to make an attempt (\textit{Enact}), potentially indicating hint abuse (one of the behaviors that signals gaming the system). 

Using the same example mentioned earlier, immediately after the student enunciated a hint: \textit{“So the molecular weight is 32.04 grams. You may want to use grams in this term, but do you want to use it in the numerator? [PROCESS]”} he continued: \textit{“Let's see what that means. I'll just put 3204 here [ENACT].”} In this example, the student read the hint and made an attempt based on it (\textit{Process} $\rightarrow$ \textit{Enact}), which may reflect part of a series of actions that constitute gaming, where the student relied on hints to derive the answer.

Additionally, we found that students were more likely to announce their errors (\textit{Errors}) before making attempts (\textit{Enact}) when gaming the system, whereas in non-gaming clips, they tended to announce their enactment followed by errors (\textit{Enact} $\rightarrow$ \textit{Errors}). The sequence \textit{Error} $\rightarrow$ \textit{Enact} may reflect systematic guessing behavior, where students repeatedly made attempts after recognizing errors, rather than making intentional attempts that subsequently led to errors (\textit{Enact} $\rightarrow$ \textit{Errors}).

\begin{table}[ht]
\centering
\renewcommand{\arraystretch}{1.2}
\caption{Strength of ordered networks. Values in \textcolor{red}{red} represent connections misaligning (compared to aligning) with the expected SRL sequence in the four-stage model \cite{winne1998srl}.}
\begin{tabular}{|l|c|c|}
\hline
\textbf{SRL Transition} & \textbf{Non-gaming} & \textbf{Gaming} \\
\hline
Process $\rightarrow$ Process & 0.04 & 0.05 \\
Plan $\rightarrow$ Plan & 0.04 & 0.02 \\
Enact $\rightarrow$ Enact & 0.13 & 0.20 \\
Errors $\rightarrow$ Errors & 0.01 & 0.02 \\
Process $\rightarrow$ Plan & 0.05 & \textcolor{red}{0.11} \\
Process $\rightarrow$ Enact & 0.08 & \textcolor{red}{0.19} \\
Process $\rightarrow$ Errors & 0.03 & 0.05 \\
Plan $\rightarrow$ Enact & 0.09 & \textcolor{red}{0.04} \\
Plan $\rightarrow$ Errors & 0.03 & 0.03 \\
Enact $\rightarrow$ Errors & 0.06 & 0.08 \\
\hline
\end{tabular}
\label{tab:lineweights}
\end{table}

\section{Discussion and Future Study}
This study explores the relationship between gaming the system (a maladaptive learning strategy and a commonly observed form of behavioral disengagement) and students' self-regulated learning (SRL) processes during interactions with an intelligent tutoring system. While gaming is typically defined and detected based on observable behaviors (e.g., systematic guessing or hint abuse), the understanding of the cognitive processes underlying such behaviors remains limited. By examining students’ think-aloud utterances through the lens of SRL, this study provides empirical evidence of the different self-regulatory processes students engage in when gaming the system. This finding deepens our understanding of the cognition surrounding gaming the system, supporting its interpretation as a maladaptive form of SRL.

In this study, we found that students’ think-aloud utterances were significantly longer during periods when gaming the system was observed. While this may seem counterintuitive (given that gaming is often associated with disengagement), it suggests that students remain cognitively active, potentially verbalizing frustration, reading and processing system feedback, or engaging in trial-and-error reasoning. This finding indicates that gaming does not simply reflect a lack of cognitive effort. To have a more nuanced understanding of how and where cognitive and behavioral engagement diverge during gaming, we further examined this phenomenon through the lens of SRL.

Our findings revealed significant differences in the engagement of SRL strategies between gaming and non-gaming periods. Specifically, students were more likely to engage in \textit{Processing Information} and \textit{Realizing Errors} during gaming clips, whereas students were more likely to engage in \textit{Planning} during non-gaming clips. These patterns suggest that gaming behavior, again, does not necessarily reflect an absence of cognitive engagement. Rather, students engage in different types of cognitive processes, specifically \textit{Processing Information} and \textit{Realizing Errors}—possibly with the aim of finding workarounds rather than genuinely learning. In contrast, the lower prevalence of \textit{Planning} during gaming suggests that students may be avoiding cognitive engagement, such as goal setting, which is a critical component of effective SRL.

Additionally, through ONA, we found during non-gaming periods, transitions among SRL strategies generally followed the expected sequence from \textit{Processing} to \textit{Planning}, then \textit{Enacting}, and finally \textit{Realizing Errors}, consistent with the four-stage SRL model \cite{winne1998srl}. This progression indicates a coherent regulatory cycle. In contrast, transitions during gaming were more disordered, for example, \textit{Planning} often preceded \textit{Processing Information}, and \textit{Realizing Errors} occurred before either \textit{Processing} or \textit{Enacting}. These reversed transitions suggest a misalignment in regulatory processes during gaming, possibly reflecting reactive or compensatory behaviors, rather than proactive learning strategies. Taken together, these findings support the interpretation that gaming may represent a maladaptive form of SRL—one in which students rely on system feedback, avoid planning, and engage in reactive rather than proactive processes.

We acknowledge the following limitations. First, the analysis relied on think-aloud protocols to infer SRL strategies. Although the think-aloud protocol is a valuable method for assessing students’ cognitive and SRL processes, it is inherently limited by what participants choose to verbalize. Some self-regulatory activities, such as monitoring or reflection, may not be captured, potentially under-representing the full scope of SRL during both gaming and non-gaming periods. Second, our data were drawn from a specific learning context, an intelligent tutoring system focused on problem-solving in stoichiometry. Thus, the findings may not generalize to other platforms, subject domains, or student populations. Different systems may afford different opportunities for both SRL and gaming behaviors, and students’ regulatory behaviors may vary. 

We also acknowledge that gaming the system can be interpreted and described differently depending on the context and goals. In certain situations, gaming may be considered an adaptive response to factors such as task completion pressures, time constraints, or system design issues. However, in the current work, we view gaming as a maladaptive strategy for the purpose of learning, as it involves bypassing meaningful engagement with the learning content in order to achieve other goals. This interpretation aligns with prior work that characterizes gaming as a maladaptive use of self-regulated learning \cite{baker2013handbook}. At the same time, we recognize that designing effective interventions to address gaming requires a more nuanced understanding — one that takes into account the underlying reasons why students engage in such behavior, including factors such as students’ motivation, time constraints, or the design of the learning platform. Future research should examine these factors alongside the conceptualization of gaming (as proposed and supported in this paper) to inform the development of more effective interventions.

\section{Conclusion}
This study provides empirical evidence demonstrating that gaming the system is not simply a form of behavioral disengagement but may involve distinct and maladaptive patterns of SRL. Through the analysis of think-aloud data, we found that students were more likely to engage in \textit{Processing Information} and \textit{Realizing Errors} during gaming clips; by contrast, they were more likely to engage in \textit{Planning} and have a more coherent and structured use of SRL during non-gaming clips. These findings support a theoretical re-framing of gaming as "self-regulated non-learning" and suggest students may still be cognitively engaged during gaming, yet misaligned with productive learning pathways.

The contributions of these results are twofold. Theoretically, they deepen our understanding of gaming by positioning it within the broader landscape of SRL, advancing thinking about gaming in terms of behavioral detection to account for the regulatory processes underlying this form of disengagement. Practically, they underscore the need for interventions that move past suppression or prevention. Rather than merely preventing, deterring, or penalizing gaming behaviors, learning systems should scaffold students’ planning and metacognitive monitoring skills that are less effective during gaming episodes. By integrating behavioral and cognitive indicators, future work can build richer models of engagement and disengagement, enabling the design of adaptive learning environments that redirect students toward more productive learning.

\begin{acks}
Carnegie Mellon University's GSA/Provost GuSH Grant funding was used to support this research.
\end{acks}

\bibliographystyle{ACM-Reference-Format}
\bibliography{main}

\end{document}